\documentclass[aps,prl,reprint,superscriptaddress]{revtex4-1}
\usepackage{blindtext}
\usepackage{graphicx}
\usepackage{amsmath,amssymb}
\usepackage{multirow}
\usepackage{url}
\usepackage{color,soul}
\usepackage[colorlinks,
            linkcolor=blue,
            anchorcolor=blue,
            citecolor=blue,
            ]{hyperref}
\usepackage{setspace}

\begin{document}

\title{Eigenfunction and eigenmode-spacing statistics in chaotic photonic crystal graphs}

\author{Shukai Ma}
\email{skma@umd.edu}
\affiliation{Quantum Materials Center, University of Maryland, College Park, Maryland 20742, USA}
\affiliation{Department of Physics, University of Maryland, College Park, Maryland 20742, USA}
\author{Thomas M. Antonsen}
\affiliation{Department of Physics, University of Maryland, College Park, Maryland 20742, USA}
\affiliation{Department of Electrical and Computer Engineering, University of Maryland, College Park, Maryland 20742-3285, USA}
\author{Steven M. Anlage}
\affiliation{Quantum Materials Center, University of Maryland, College Park, Maryland 20742, USA}
\affiliation{Department of Physics, University of Maryland, College Park, Maryland 20742, USA}
\affiliation{Department of Electrical and Computer Engineering, University of Maryland, College Park, Maryland 20742-3285, USA}

\begin{abstract}

The statistical properties of wave chaotic systems {of varying dimensionalities and realizations} have been studied extensively.
{These systems are commonly characterized by the statistics of the eigenmode-spacings and the statistics of the eigenfunctions.} 
Here, we propose photonic crystal (PC) defect waveguide graphs as a new physical setting for chaotic graph studies.
{Photonic crystal waveguides possess a dispersion relation for the propagating modes which is engineerable.
Graphs constructed by joining these waveguides possess junctions and bends with distinct scattering properties.}
{We present numerically determined statistical properties of} an ensemble of such PC-graphs {including} both eigenfunction amplitude and eigenmode-spacing studies.
{Our proposed system is compatible with silicon nanophotonic technology and opens chaotic graph studies to a new community of researchers.}

\end{abstract}

\maketitle

\begin{spacing}{1}

\section{I. Introduction}

Wave-chaotic phenomena have been studied in various complex scattering systems, ranging from 1D graphs \cite{Hul2004,Martinez-Arguello2018,Rehemanjiang2020,Lu2020,Chen2020}, 2D billiards \cite{Wu1998,Dembowski2003,Dembowski2005,Hemmady2005,Kuhl2008,Shinohara2010,Dietz2019} to 3D enclosures \cite{Hemmady2012,Xiao2018,Ma2019,Ma2020,Ma2020b,Frazier2020}.
The statistical properties of many system quantities, such as the closed system eigenvalues and the open system scattering/impedance matrices, exhibit universal characteristics, which only depend on general symmetries (e.g., time-reversal, symplectic) and the degree of system loss.
{Random Matrix Theory (RMT) has enjoyed great success in describing the energy level spacing statistics of large nuclei} \cite{Mehta2004,Dietz2017prl}.
{The statistics of complex systems that show time-reversal invariance (TRI) are described by the Gaussian orthogonal ensemble (GOE) of random matrices, and the statistics of systems showing broken time-reversal invariance are described by the Gaussian unitary ensemble (GUE), and the statistics of systems with TRI and antiunitary symmetry show Gaussian symplectic ensemble (GSE) statistics } \cite{So1995,Hemmady2005,Ma2020}.
{It was later conjectured by Bohigas, Giannoni, and Schmit that any system with chaotic dynamics in the classical limit would also have semi-classical wave properties whose statistics are governed by RMT associated with one of these three ensembles} \cite{Bohigas1984}. 
{In practice, identifying RMT statistical properties in experimental data is difficult due to the presence of non-universal features that obscure the universal fluctuations}.
The Random Coupling Model (RCM) has found great success in characterizing the statistical properties of a variety of experimental systems by removing the non-universal effects induced by port coupling and short-orbit effects \cite{Hemmady2005,Zheng2006,Hart2009,Yeh,Hemmady2012,Li2015,Auregan2016,Xiao2018,Zhou2019,Ma2020,Ma2020b}.

Chaotic microwave graphs support complex scattering phenomena despite their relatively simple structure \cite{Hul2004,Martinez-Arguello2018,Rehemanjiang2020,Lu2020,Chen2020,Zhang2022}.
{We refer loosely to a chaotic graph as one in which there are multiple paths of differing lengths from one node to another, such that waves traveling along these paths interfere when arriving at nodes.}
A microwave graph structure can be realized with coaxial cables as graph bonds connected at T-junctions.
The graph architecture allows for various useful circuit components (such as variable phase shifters and attenuators, as well as microwave circulators) to be incorporated into the structure \cite{Rehemanjiang2020,Lu2020,Chen2020}.
Recent studies show that non-universal statistical features exist in chaotic graph systems, and these are hypothesized to be caused by the non-zero reflection at the graph vertices.
These reflections create trapped modes that impact the spectral statistics of the graph \cite{Biaous2016,Dietz2017,Lu2020,Che2021}.

\begin{figure*}
\centering
\includegraphics[width=1\textwidth]{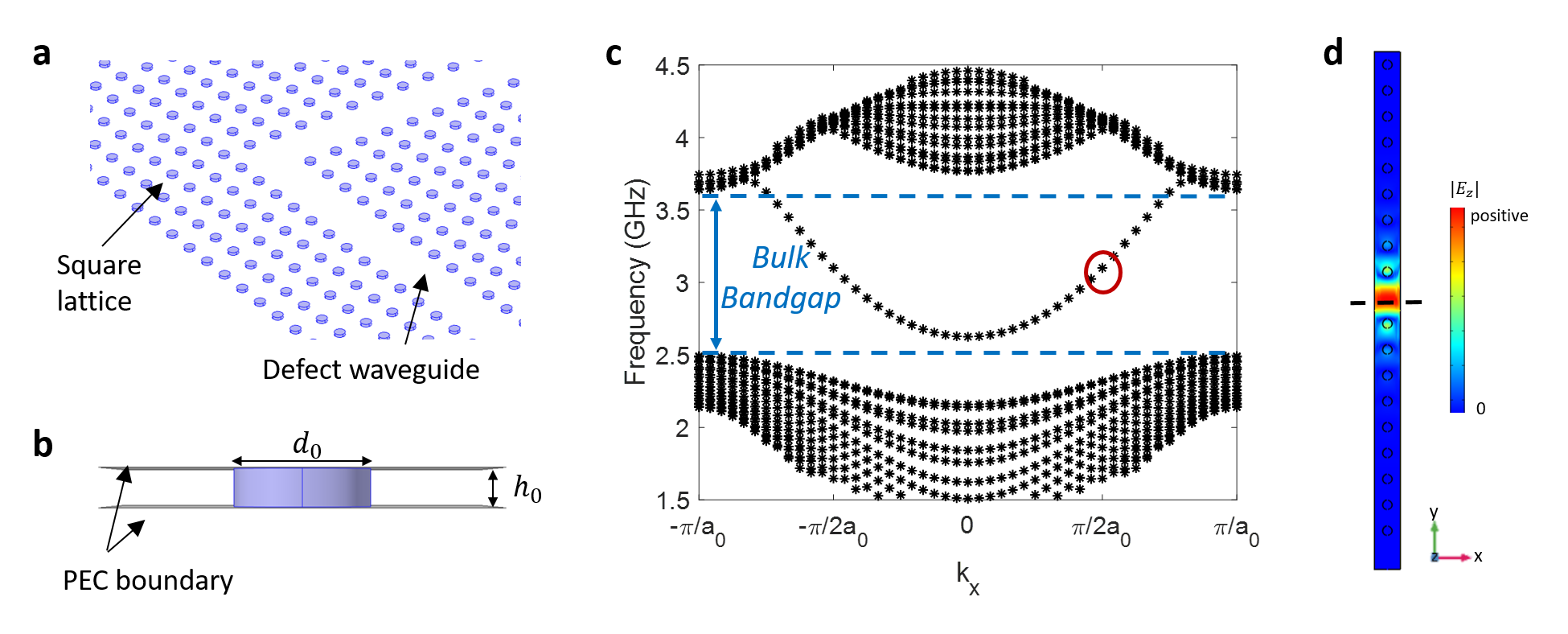}
\caption{\label{fig:fig1} 
\textbf{a.} The open-plate view of the photonic crystal lattice with an L-shaped defect waveguide region.
\textbf{b.} The side view of the PC unit cell. A dielectric rod is sandwiched between two PEC (perfect electric conductor) surfaces. 
The quantities $h_0$ and $d_0$ are the height and diameter of the rod.
\textbf{c.} The photonic band structure from a supercell defect waveguide simulation. The waveguide modes appear in the bulk bandgap region ({delineated by the blue dashed lines}).
\textbf{d.} The E-field profile of one of the waveguide mode solutions (the red circle in \textbf{c}). The black dashed line marks the center of the waveguide.
}
\end{figure*}

Here, we introduce an alternative type of chaotic graph built with photonic crystal (PC) waveguides.
{Photonic crystals find extensive use in semiconductor-based nanophotonic technology, where they act as low-loss waveguides for visible and infrared light} \cite{Johnson2001}.
The PC graph bonds are realized with defect waveguides, and the nodes are formed by the waveguide junctions.
{As TRI is preserved and spin-1/2 behavior is absent, chaotic PC-graph systems are expected to fall into the GOE universality class of RMT.}
With numerical simulation tools, we conduct a series of statistical tests of chaotic photonic crystal graph systems including both eigenvalue and eigenfunction studies of closed graphs.

{The unique properties of PC structures for quantum chaos studies are as follows.
First, the bond lengths of PC-graphs can be altered by means of lithographic fabrication} \cite{Johnson2001}. 
{Second, the scattering properties of PC-graph nodes can be engineered by changing the rod properties and geometry.  
Moreover, PC defect waveguides and cavities are technologies that are widely used in the field of integrated nanophotonics. 
To our knowledge, PC waveguide graphs have not been previously utilized for quantum graph studies.}

The paper is organized as follows.
In Section II, we introduce the design details of the proposed PC graph structure as well as its numerical implementation methods. 
We present the chaotic closed-graph mode-spacing study in section III, and focus on the discussion of closed-graph eigenfunction studies in section IV.
Different methods of conducting eigenfunction statistical studies are also discussed. 
We summarize the paper and discuss the future applications of the proposed PC graphs in section V.

\section{II. Photonic crystal graph}

A photonic crystal system consists of a {regular} lattice of artificial atoms (or scatterers) whose spacing is comparable to the operating wavelength \cite{Joannopoulos1997,Joannopoulos2008}.
The material properties and the geometrical details of the atoms are carefully designed in order to achieve a specific functionality.
A PC system is usually constructed as a 2D planar structure, which {makes it} especially good for lithographic fabrication and {photonics} applications.
Importantly, a 2D PC can show a complete bulk bandgap in its electromagnetic excitation spectrum \cite{Joannopoulos1997,Joannopoulos2008}.
In PC-based devices, waveguides and cavities can be constructed {for example} by making air defects (removing a certain number of the atoms) in the original lattice.
This creates guided propagating modes in the bulk bandgap, which ensures that the modes are confined to the defect region.
{A variety of defect waveguides can be realized by changing the atom properties \cite{Joannopoulos1997}.}
Recent photonic topological insulator (PTI) studies present an alternative form of PC waveguide using the interface between two different topological domains \cite{Xiao2016,Lai2016,Ma2019pti,Ma2020pti,Liu2021}. 
{Such PTI-based waveguides can also be utilized as the bonds for realizing graphs. 
The interesting properties of the topologically protected modes may further enhance the quantum graph studies.}
Here, we will utilize the defect waveguide modes to build chaotic graph structures. 
{Note that these PC defect waveguide graphs are distinctly different from photonic crystal slabs and billiards \cite{Dietz2019,Maimaiti2020,Zhang2021}, and from coupled resonant dielectric cylinders \cite{Rehemanjiang2020,Reisner2021} studied previously.}

\begin{figure*}
\centering
\includegraphics[width=0.95\textwidth]{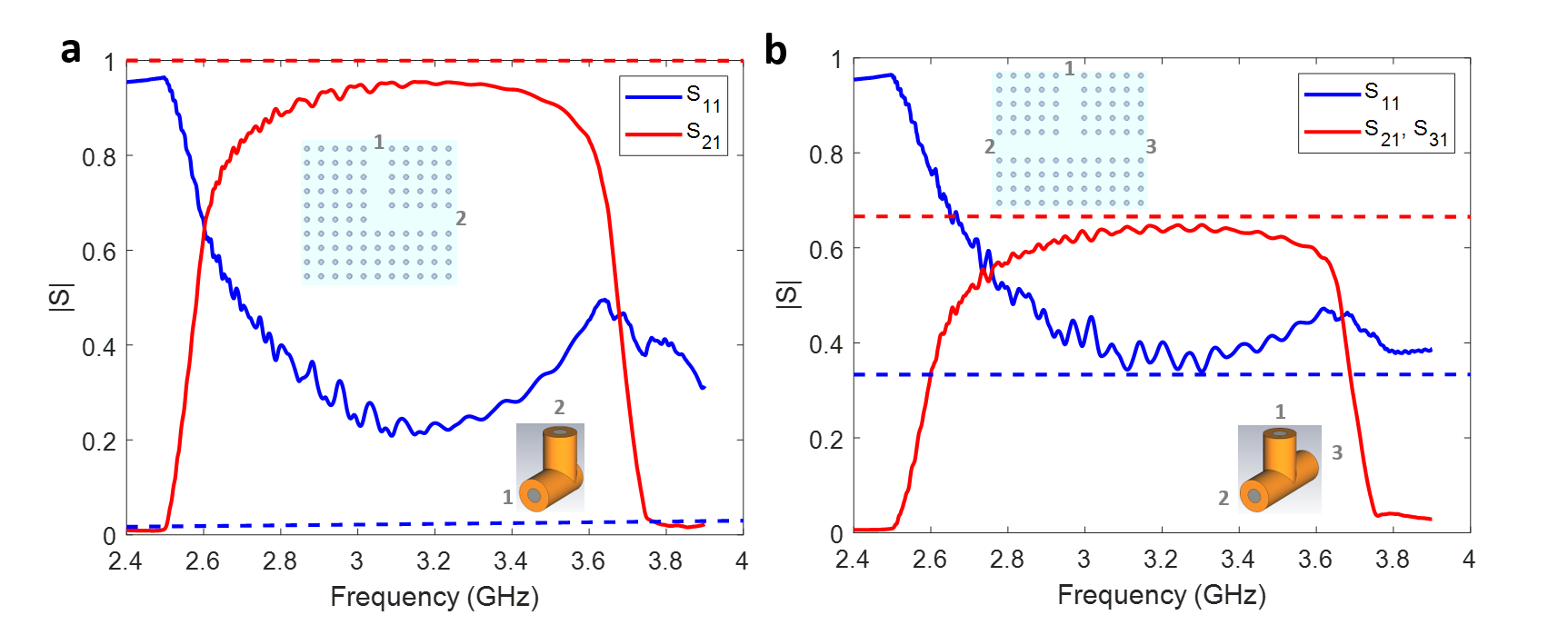}
\caption{\label{fig:fig_cst} 
\textbf{a.} The simulated S-parameters of the PC-graph right-angle junction (solid), as well as the S-parameter from a $3.5mm$ uniform coaxial cable 2-port right-angle connector (dashed), which serves as a reference.
{Inset: the schematics of the PC right-angle junction and the 2-port coaxial cable junction.}
\textbf{b.} The simulated S-parameters of the PC-graph T-junction (solid) and the S-parameters from a coaxial cable 3-port `T' connector (dashed).
{Inset: schematics of the PC T-junction and the 3-port coaxial cable junction.}
All PC-junction S-parameters are {frequency averaged with a 500MHz window to remove the effect of spurious reflections.}
}
\end{figure*}

The construction of the chaotic PC graph starts by building a square lattice with identical dielectric rods (the blue lattice in Fig. \ref{fig:fig1}(a)).
The lattice constant is $a_0 = 36.8mm$.
The dielectric rod lattice is located between two metallic surfaces and installed in a vacuum background.
The defect waveguide is created by simply removing one row or column of the dielectric rods.
The detailed shape of the dielectric rod is shown in Fig. \ref{fig:fig1}(b).
The diameter of the dielectric rod is $d_0 = 0.36a_0 = 13.2mm$ and the height is $h_0 = 0.1a_0 = 3.68mm$.
Because the PC lattice is thin in the vertical direction ($z$-direction), the waveguide modes {considered here} are polarized such that $E_z \neq 0$, $E_{x,y} = 0$, $H_z = 0$ and $H_{x,y} \neq 0$ (TM with respect to $z$), {and these are the only modes that propagate for frequencies $f < 12$ GHz, well above the band gap of interest here}.
The relative permitivity and permeability of the dielectric rods are $\epsilon_r = 11.56$ and $\mu_r = 1$.
We realize the proposed PC structure numerically with COMSOL Multiphysics Software.
The presence of the waveguide mode is clearly demonstrated by the supercell photonic band structure (PBS) simulation (Fig. \ref{fig:fig1}(c)).
The supercell simulation model consists of a single column of PC lattice with Floquet periodic boundaries on the two long sides.
{That is, fields on opposite long sides differ by a phase factor $\theta = k_x a_0$.}
Fields on the opposite short sides are assigned totally absorbing boundary conditions.
We remove the center rod to create the defect waveguide region.
The PBS simulation is conducted by computing the system eigenmodes while varying the wavenumber $k_x$ in the range $[-\pi/a_0, \pi/a_0]$.
As shown in Fig. \ref{fig:fig1}(c), the defect waveguide modes have emerged inside of the bulk bandgap region from $2.5 \sim 3.6$GHz.
{The eigenfrequencies in the band-gap region are essentially real, as the absorbing boundaries are well separated from the defect creating the waveguide.}

{We see from Fig.} \ref{fig:fig1}(c) {that the waveguide modes in the PC have a distinct dispersion relation (frequency versus $k_x$) that differs from that of a transmission line, $\omega=k_x c$. 
The dispersion relation is parabolic at the lower band-edge, as in a metallic waveguide.  
However, the group velocity vanishes at the upper band-edge, in contrast with that of a metallic waveguide for which there is no upper band-edge, rather just a boundary where new propagating modes are possible} \cite{Zhang2022}.
In Fig. \ref{fig:fig1}(d), the $|E_z|$ profile of the entire simulation domain shows clearly that the mode solution is indeed a guided wave because its amplitude is highly localized in the defect region.

{In addition to the propagation on the graph bonds, it is also important to characterize the scattering properties of graph nodes} \cite{Bittner2013}.
The PC graph structure is realized by connecting multiple straight defect waveguides with both right-angle and T-shaped junctions.
We have characterized the scattering matrix of both types of junctions with COMSOL and CST Microwave STUDIO.
For the right-angle junctions, non-zero transmission is found only in the bulk bandgap region.
{As shown in Fig. \ref{fig:fig_cst}(a), the transmission $|S_{21}|$ and the reflection $|S_{11}|$ for right angle bends vary systematically as functions of frequency.
Thus, right-angle bends are vertices with a non-trivial scattering matrix in a graph.}
The PC-graph T-junction also presents a complex scattering profile over the entire bandgap region.
As shown in Fig. \ref{fig:fig_cst}(b), the magnitude of the reflection (transmission) coefficient deviates systematically from 1/3 (2/3) as a function of frequency.

{It should be noted that for the PC graph junctions the sum of the squares of the displayed scattering coefficients is close to unity.  
This implies that at the frequencies of interest the photonic waveguides are operating in a single transverse mode.}
{For an ideal transmission line there is no reflection at a bend. 
Also, at an ideal T-junction the reflection coefficient $|S_{11}|=|-1/3|$, and the transmission coefficient is $|S_{12}|=|2/3|$. 
The ideal scattering parameters apply to transmission lines whose transverse dimensions are much smaller than an axial wavelength (equivalently the frequency is well below the first cut-off frequency of non-TEM modes.).  
The calculated scattering properties of an air-filled coaxial ($r_o = 2$ mm, $r_i = 0.86$ mm) cable bend and T-junction are shown as well in Fig.} \ref{fig:fig_cst}, {and, by contrast, they show minimal variation in the frequency range of interest.}
{The cut-off frequency of the $TE_{11}$ mode in this cable is 33.34 GHz, well above the operating frequency. 
Consequently, the scattering parameters are close to the ideal values.  
Graphs constructed of metallic waveguides, as opposed to transmission lines, can be expected to have scattering parameters with characteristics similar to PC graphs for frequencies between the first and second lowest cut-off frequencies where single mode operation is possible} \cite{Zhang2022}.
These nontrivial junction scattering parameters are a second unique feature of PC graphs, and {can be expected to influence the properties of} the graph eigenmodes.
We note that alternative methods of making waveguide bends, or connecting waveguides, can be applied, for example by removing or adding dielectric rods at the right-angle turn, in order to tune the transmission property of the waveguide joints \cite{Mekis1996}.
{One may adjust the degree of wave localization of graph nodes by engineering the scattering properties of the PC waveguide junctions.}
The engineering of the node reflection and transmission properties is beyond the scope of this paper.

\begin{figure*}
\centering
\includegraphics[width=1\textwidth]{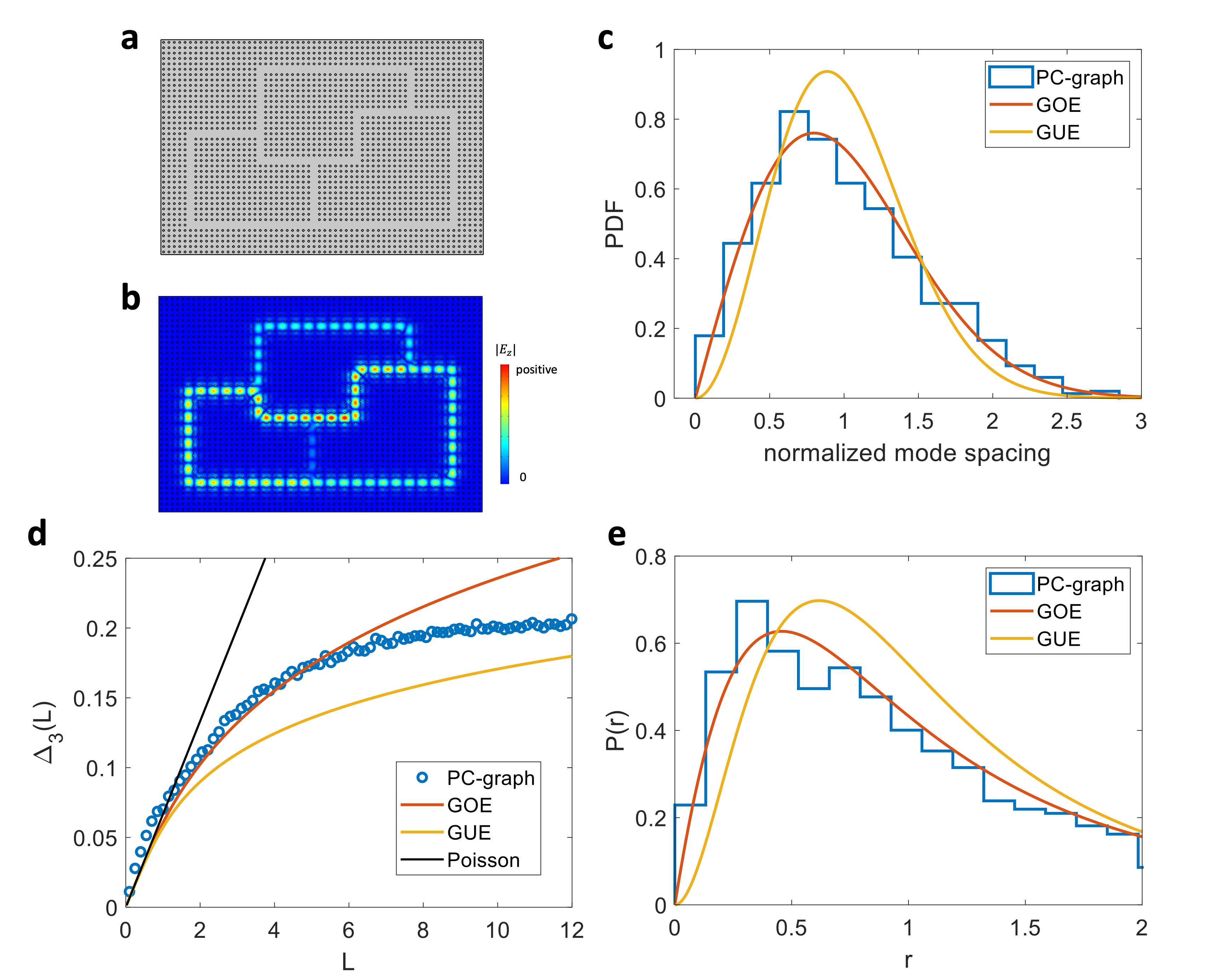}
\caption{\label{fig:fig2} 
\textbf{a.} Schematic diagram of the PC waveguide graph system. The graph bonds are shown as clear gray channels, and the rectangular lattice is shown as a lattice of dielectric rods (black circles).
\textbf{b.} The simulated $|E_z|$ profile of the graph system eigenmode at 2.8GHz.
\textbf{c} shows the statistics of the normalized mode spacing from graph simulation (histogram). 
The theoretical predictions for the GOE and GUE systems are shown in red and yellow curves.
\textbf{d} shows the spectral rigidity $\Delta_3(L)$ of the normalized spectrum from graph simulation (blue circles). The theoretical predictions for the GOE, GUE, and the Poisson systems are shown in red, yellow, and black curves.
\textbf{e} shows the statistics of the consecutive mode spacing ratio $r$ (histogram) and the theoretical predictions for the GOE and GUE systems. }
\end{figure*}

We simulate the closed PC graphs with the COMSOL eigenvalue solver.
The graph topology is that of a flattened tetrahedral graph having $14$ straight segments and $13$ junctions (including both 3-way junctions and right-angle junctions).
The four exterior faces of the {parallel-plate PC structure (see Fig. \ref{fig:fig2}(a))} are assigned totally absorbing boundary conditions.
The total length of the simulated graph is on the scale of $\sim 9$ m, which hosts about 80 eigenmodes (within the bulk PC bandgap) in a typical realization.
We note that one is able to decrease the mode-spacing by enlarging the size of the PC graph, and we choose the current system scale due to limited computational power.
We have created a statistical ensemble of chaotic PC graphs by changing the length of the bonds for a given graph topology.

{A particular eigenmode solution of a graph} {(shown in Fig. \ref{fig:fig2}(a))} is shown in Fig. \ref{fig:fig2}(b).
It is clear that the graph mode is localized to the defect waveguide region and displays a longitudinal sinusoidal standing wave pattern. 
The mode amplitude on a single bond is uniform but varies between different bonds (shown by the color differences), where bonds are defined as straight waveguide regions between 3-way junctions and right-angle bends.

\section{III. Eigenmode Spacing Statistical Analysis}

{In order to characterize the statistical properties of PC defect waveguide graphs,} we start by conducting the nearest neighbor mode spacing analysis of the proposed PC graphs.
An ensemble of 10 different graph realizations is studied numerically, and we obtained $\sim 800$ eigenfrequency values from the ensemble.
The graph topology ranges from $13\sim16$ straight segments and $11\sim14$ junctions.
We note that this ensemble of graphs is conducted with the eigenvalue simulation in COMSOL. 
{The simulation results are used for mode-spacing studies here, as well as the eigenfunction statistics in Sec. IV.}
{The model is composed of lossless dielectric structures.  
However, the presence of radiating boundary conditions on the perimeter of the finite-size structure shown in} Fig. \ref{fig:fig2}(a) {will give rise to a finite loss.  
The quality factors of the simulated graph modes are on the order of $10^5$ to $10^7$, allowing for clear identification of all eigenfrequencies.}
The graph topology is kept fixed in the eigenfunction studies discussed below.
Because the topology and the total length $L_{tot}$ of each graph realization are different, we normalized the system eigenmode solutions using the following method.
We considered 10 different graphs. 
For each graph, we calculated eigenfrequencies in the range 2.8 GHz to 3.6 GHz and placed the eigenfrequencies in ascending order.  
The eigenfrequencies were then converted to eigen-wavenumbers ($k_i$, $i$ is the mode index) using the dispersion relation for the PC-waveguide PBS shown in Fig. \ref{fig:fig1}(c).  
{This results in a tabulation of consecutive wave-numbers. }
We then fit the mode-number index to the wave-number using a quadratic fitting function $n(k) = c_2 k^2 + c_1 k + c_0$, where $c_{0,1,2}$ are fitting parameters and $n(k)$ is the total number of modes below $k$ \cite{Rehemanjiang2016}.
For a typical graph we found $c_2 = 0.0054 m^2, c_1 = 2.21 m$ and $c_0 =-55.28$.  The small value of $c_2$ indicates the $n(k)$ is nearly linear in $k$ as expected for quasi-one-dimensional structures.  
We then {evaluate} the fitting functions at the calculated $k$-values for each graph, creating a list of normalized eigenvalues $e_i = n_{fit}(k_i)$.  The nearest neighbor spacings {are} finally computed as $s_i = e_{i+1} - e_i$.

The distribution of the normalized nearest-neighbor mode spacing values of the entire ensemble is shown as the histogram in Fig. \ref{fig:fig2}(c).
{We have included the approximate theoretical predictions of the mode-spacing statistics for the GOE and GUE systems in the figure.}
These theoretical predictions are based on RMT \cite{Berry1981,Kottos1997,Kottos1999}, which are methods of understanding the universal statistical properties of wave-chaotic systems \cite{Casati1980,Bohigas1984}.
As shown in Fig. \ref{fig:fig2}(c), the distribution of the normalized mode-spacing matches reasonably well with the theoretical prediction for the GOE universality class.
Good agreement between the graph nearest neighbor mode-spacing statistics and the RMT theoretical prediction is also reported in various works on graphs, although long-range statistical quantities tend to show non-universal behavior due to the trapped-mode issue mentioned above \cite{Biaous2016,Dietz2017,Lu2020,Che2021}.

{To test the long-range statistical properties of the eigenmodes, we compute the spectral rigidity of the graph spectrum $\Delta_3(L)$, where $L$ measures the length of a segment in the normalized spectrum.}
Our procedure is as follows.  
The quantity $\Delta_3(e, L)$ representing the deviation of the spectral staircase from a straight line, is evaluated for each value of $L$ by 
\begin{equation}
    \Delta_3(e, L) = \frac{1}{L} min_{a, b} \int_e^{e+L} [N(x) - ax - b]^2dx,
\end{equation}
where $N(e) = n(k)$ and $e$ is chosen at random.
{We use 100 values of e for each graph.  Then we average the values of e over 10 graph realizations} \cite{haq1982fluctuation}.
The resulting values of $\Delta_3(e, L)$ are then plotted in Fig. \ref{fig:fig2}(d) along with theoretical predictions based on GOE, and GUE random matrices \cite{Che2021}, and the predictions of Poisson statistics \cite{berry1985semiclassical} (appropriate for classically integrable systems).

{The $\Delta_3(L)$ curve in} Fig. \ref{fig:fig2}(d) {is linear in $L$ for small values of $L$.  
As shown, systems of all symmetries (Poisson, GOE, GUE) show a linear-in-$L$ dependence for $L<1$} \cite{berry1985semiclassical}, {and the data is consistent with this trend.  
Chaotic systems are then  expected to show a logarithmic increase in $\Delta_3(L)$} \cite{berry1985semiclassical}, {and this is evident in the data up to approximately $L=5$.  
It is also predicted that $\Delta_3(L)$ will saturate due to the presence of short orbits in any experimental realization of a wave chaotic system} \cite{berry1985semiclassical}, {and this is also seen in the data.}
Previous studies find that the long-range statistics of quantum graphs usually show some degree of agreement with the GOE prediction in the range of roughly $L \in [2, 5]$, along with deviations beyond this range \cite{Hul2004, Biaous2016, Dietz2017}.
We observe in Fig. \ref{fig:fig2}(d) a deviation from the GOE RMT predictions, especially for larger $L$, similar to cable graph studies in Ref. \cite{Hul2004}.
{Because the operating bandwidth of the proposed PC-graph system is relatively narrow} (Fig. \ref{fig:fig1} (c)), {we believe the long-range statistics may also be affected by finite-size effects because the number of available eigenmodes is bounded above and below by the edge of the PC bandgap.}
{The solution is to increase the graph size substantially, but this is beyond our computational resources.}
{In summary, the $\Delta_3(L)$ behavior is very similar to that observed in wave chaotic systems of higher dimensionality.}

\begin{table}
\begin{ruledtabular}
\begin{tabular}{l l l l l}
  & PC-Graph & GOE & GUE & GSE\\
  \hline
$\left< r \right>$ & 1.89 & 1.75 & 1.37 & 1.18\\
$\left< \Tilde{r} \right>$ & 0.51 & 0.54 & 0.60 & 0.68\\
\end{tabular}
\end{ruledtabular}
\caption{\label{tab:catalog} The summarized consecutive mode spacing ratios of the photonic crystal graph (PC-Graph) system, and the theoretical predictions for the GOE, GUE, and GSE systems \cite{Atas2013}. }
\end{table}

In addition to the mode-spacing distribution test, we note that the method of consecutive mode spacing ratios $r_i = \frac{s_i}{s_{i-1}}$ and $\Tilde{r}_i = min \left( r_i, \frac{1}{r_i} \right)$ can also be adopted \cite{Atas2013}.
{Here the ratio statistic has the feature that it is a measure of the correlation between adjacent spacings.
One advantage of this type of statistical study is that it does not require unfolding the spectrum} \cite{Atas2013}.
{However, here we use the unfolded spectrum $s_i$ to compute the spacing ratios.}
For the PC graphs, the averaged values of $\left< r \right> = 1.89$ and $\left< \Tilde{r} \right> = 0.51$, are closer to the GOE theoretical predictions than the GUE prediction (Table. \ref{tab:catalog}).
We note that the high value of $\left< r \right>$ in the PC-graph is related to the correlation between consecutive level-spacings.
We further present the distribution of mode-spacing ratios $r$ of the PC graphs and corresponding theoretical predictions of the GOE and GUE systems in Fig. \ref{fig:fig2}(e).
The statistics of $r$ match reasonably well with the GOE prediction.

\section{IV. Eigenfunction Analysis}

We next study the statistics of the {closed} PC graph eigenfunctions.
The eigenfunction statistics have been studied experimentally in 2D chaotic systems by probing the electromagnetic (EM) standing wave field inside a microwave cavity \cite{Kudrolli1995,Wu1998,Gokirmak1998,Chung2000,Kuhl2000,Hlushchuk2001}.
In those studies, the experimental probability amplitude distribution and two-point correlation function agree well with the {random plane wave conjecture}.
Here, the wave properties of the entire {closed} PC graph (nodes and bonds) can be faithfully simulated.
We employ the same eigenvalue simulation model used in the mode-spacing studies above.
For a graph system, the telegrapher's equation is formally equivalent to the 1D Schr\"{o}dinger equation, where the wavefunction $\psi$ is represented by the wave voltage.
For thin parallel plate waveguides, the wave voltage difference between the top and bottom metallic plates is represented by the $E_z$ value at the middle cutting plane (at the height of $z = h_0/2$).
Because the PC graph bonds have a finite width, the bond eigenfunction will be evaluated along a 1D line at the center of the waveguide (shown as the black dashed line in Fig. \ref{fig:fig1}(d)).
We next examine two PC graph eigenfunction characterization methods.

\begin{figure*}
\centering
\includegraphics[width=0.9\textwidth]{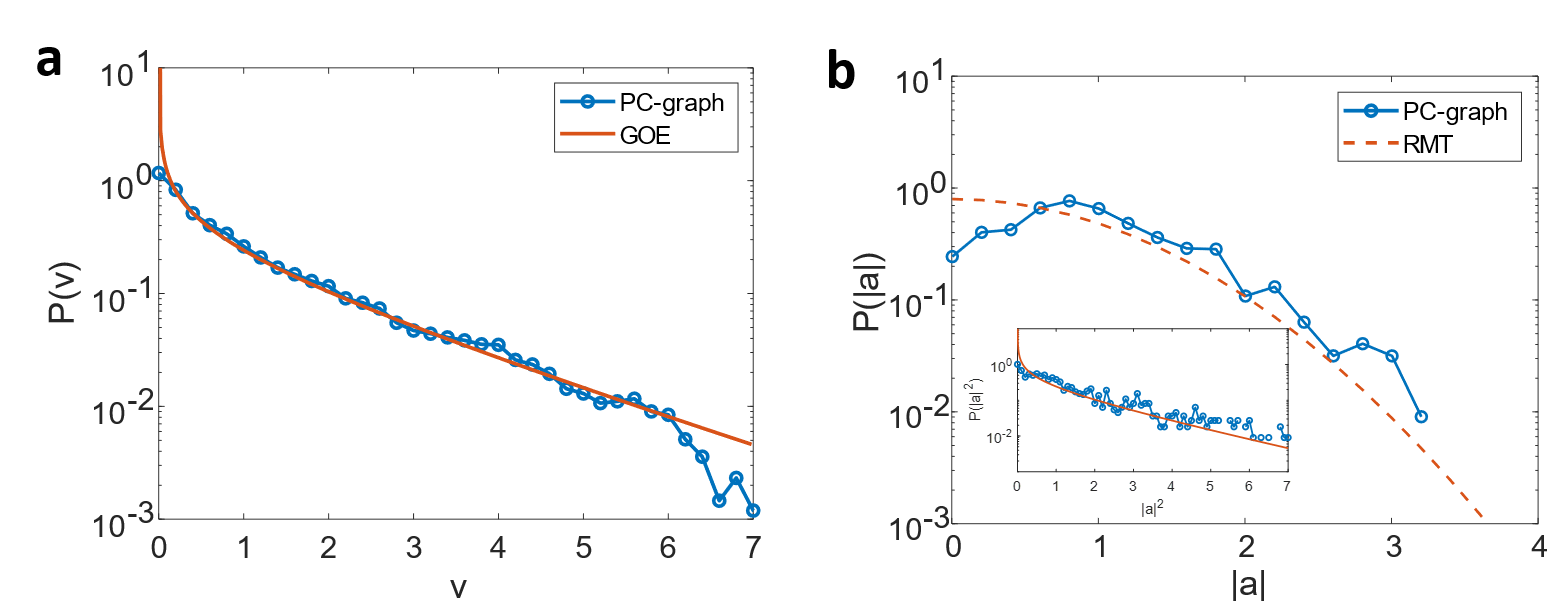}
\caption{\label{fig:fig3} 
\textbf{a.} Probability amplitude distribution of photonic crystal defect waveguide graph eigenmodes values ($v$) obtained from the grid-wise method (symbols connected by blue line). 
The theoretical prediction for systems with GOE statistics is in red.
\textbf{b.} shows the statistics of the normalized bond amplitude value $|a|$ (symbols connected by blue line) and the Gaussian distribution (red dashed).
{The inset shows the distribution of the quantity $|a|^2$ and the Porter-Thomas theoretical prediction for the GOE system.}
}
\end{figure*}

\textit{Method I: grid-wise representation}.
For each eigenmode of each graph realization, we will use the entire set of the graph bond $|E_z|^2$ vs. longitudinal position along the center-line of the waveguide data points to represent the eigenfunction.
The name `grid-wise' comes from the granular nature of this method, where the total number of eigenfunction data points is inversely proportional to the computational grid size.
We study the wavefunction statistics of the PC graph by computing the distribution of the normalized probability density $v$, which is defined as the square of the eigenfunction values
\begin{equation}
    v_j = |\psi(r_j)|^2 = \frac{|E_z(r_j)|^2 \cdot L_{tot}}{\sum_j |E_z(r_j)|^2 \cdot \Delta L_j}, 
\end{equation}
where $L_{tot}$ is the total length of the graph, $r_j$ and $\Delta L_j$ are the location and the grid size of the $j'th$ grid point, $|E_z(r_j)|$ is the z-directed electric field magnitude at the $j'th$ grid point, and the summation in the denominator runs over every graph grid point.
In our simulation, the grid size $\Delta L_j \sim 0.05 \lambda_{op}$ where $\lambda_{op} = 10cm$ is the operating wavelength at $3GHz$.
The distribution of the probability density values is computed using the data from {all simulated eigenmode solutions (48 to be specific)} from a single realization, and the results are shown in Fig. \ref{fig:fig3}(a), and discussed below.

\textit{Method II: bond-value representation}.
For each graph realization, the eigenfunction of a mode is represented by a set of `bond-values' $E_z(b_m)$ which is defined as the amplitude of the standing-wave wavefunction on the graph bond $b_m$. 
The quantity $m$ is the index of the bond and runs from 1 to 14.
The standing wave on the bond is made up of two counter propagation waves $\psi(x) = a_m \, e^{ikx} + a_m^* \, e^{-ikx}$, where $a_m$ is the wavefunction amplitude at bond $b_m$ and $x$ is the distance from a vertex along the bond.
{We first conduct a sine-fit of the raw $E_z(x)$ values on each bond, which yields the amplitude, and the value of $|a_m|$ is obtained as 1/2 of the amplitude value.}
The normalization process follows the same method as in Ref. \cite{Kaplan2001} which ensures that $\sum_m L(b_m) |a_m|^2 = L_{tot}$ where $L(b_m)$ is the length of bond $b_m$.
Here the distribution of the probability density values is computed over $14 \times 78$ data points, where 14 is the number of the bonds and 78 is the number of eigenmode solutions from one graph configuration, and the results for $P(|a|)$ are shown in Fig. \ref{fig:fig3}(b).

{The random plane wave hypothesis underlying chaotic eigenfunction treatments asserts that the eigenfunctions in a 2D or 3D cavity can be thought of as superpositions of plane waves with random directions and random phases.  
Consequently, the value of an eigenfunction at any point is a sum of many random waves and is thus a zero mean Gaussian random variable} \cite{Berry1981}. 
{For systems with TRS the plane waves come in counter-propagating pairs, and the wave function is real. 
For systems with TRS broken, wave directions are completely random, and the wave function is complex with independent real and imaginary parts.  
For the PC graph the situation is changed in that at any point in the graph there are only two counter-propagating waves of equal amplitude. 
The amplitudes of these waves, $|a_m|$ will be different on the different bonds leading to statistical variations.  }

\begin{figure*}
\centering
\includegraphics[width=0.9\textwidth]{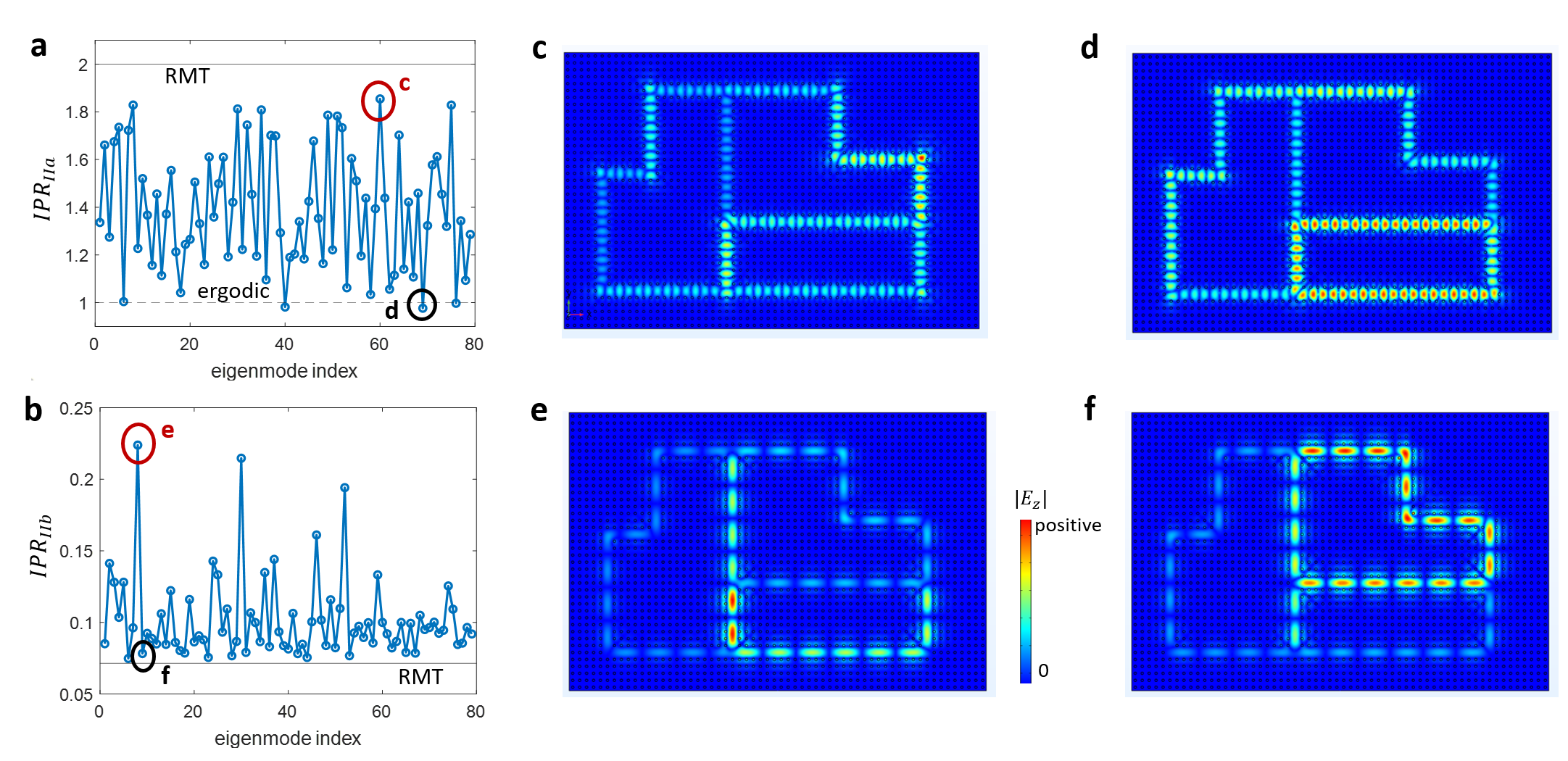}
\caption{\label{fig:fig4} 
\textbf{a.} The inverse participation ratio (IPR) of graph eigenmodes obtained from the bond-value method, computed based on method IIa. The simulated $|E_z|$ profile of the red (black) circled mode in \textbf{a} is shown in \textbf{c} (\textbf{d}).
\textbf{b.} The inverse participation ratio (IPR) of graph eigenmodes is computed based on method IIb.
The simulated $|E_z|$ profile of the red (black) circled mode in \textbf{b} is shown in \textbf{e} (\textbf{f}).
}
\end{figure*}

We first discuss the eigenfunction statistics obtained using the grid-wise method in Fig. \ref{fig:fig3}(a).
{For waves in a 2D or 3D cavity the distribution of values of $v$ based on the random plane wave hypothesis is $P(v) = (2\pi v)^{-1/2} e^{-v/2}$.}
This is shown as a solid red line in Fig. \ref{fig:fig3}(a).
{We find that the simulated PC graph result matches to some degree the GOE prediction in that for $v > 1$ the curve approaches a straight line on a log-linear plot, and for $v$ small there is an indication of square root singularity.  
We notice that the mismatch between the graph data and the GOE prediction exists at the small and large probability density values, indicating an absence of both very small and very large $v$ values. 
One possibility is that this method tends to over-count the appearance of medium-sized eigenfunction values (similar to the systematic errors experienced in} Ref. \cite{Wu1998}).
It may also indicate that the data simply does not match the GOE prediction.
In addition to the discrete eigenfunction imaging method, we note that the eigenfunction statistics may also be tested through resonance width distributions in transmission measurements (Porter-Thomas statistics) \cite{Alt1995, Dembowski2005}.

{Here we now use \textit{Method II} to study the distribution of normalized bond values, $|a|$ that are displayed in} Fig. \ref{fig:fig3}(b).  
{Also plotted in} Fig. \ref{fig:fig3}(b) {is a Gaussian distribution, which would be characteristic of an eigenfunction of a GOE random matrix.}
However, a clear deviation from Gaussian statistics is seen for low amplitudes, similar to the deviation observed with \textit{Method I}.
{The inset to} Fig. \ref{fig:fig3}(b) {shows a plot of $P(|a|^2)$ compared to the Porter-Thomas distribution expected for GOE systems.  
There is an agreement between the two except for deviations between data and the model distribution at small values of $|a|^2$,  similar to that seen in} Fig. \ref{fig:fig3}(a).
Together, these results suggest that the wavefunction statistics of this simple graph are not fully consistent with the random plane wave hypothesis.

We next present the inverse participation ratio (IPR) computation based on the bond-value method (\textit{Method II}) in Fig. \ref{fig:fig4}.
IPR is a measure of the degree of localization for a wave function \cite{Kaplan2001, Hul2009}, and can be used to quantify the degree of deviation of wavefunction statistics from the random plane wave hypothesis.
Based on its IPR value, the eigenfunction behavior varies between two limits, namely the maximum ergodic limit where the wave function occupies each graph bond with equal chance, and the maximum localization limit where the eigenmode is confined to only one bond. 
RMT predicts an IPR value by assuming Gaussian random fluctuation of the eigenfunctions.
Two different IPR definitions are tested here.
Method IIa follows the definition in Ref. \cite{Kaplan2001}, where the IPR value for each graph mode is evaluated using the formula $IPR_{IIa} = \langle \left| a_m \right|^4 \rangle$.
As shown in Fig. \ref{fig:fig4}(a), the IPR values of the photonic crystal graph modes vary erratically but lie between the maximum ergodic limit ($IPR_{IIa}$ = 1) and the RMT prediction limit ($IPR_{IIa}$ = 2) \cite{Kaplan2001}.
In the maximum localization limit the $IPR_{IIa}=B$, where $B=14$ is the number of graph bonds, and the results are far from this limit.
Method IIb follows the definition in Ref. \cite{Hul2009} where $IPR_{IIb} = \sum_m \left| \Tilde{a}_m \right|^4 / \left[ \sum_m \left| \Tilde{a}_m \right|^2 \right]^2$. 
The quantity $\Tilde{a}_m$ is the un-normalized bond wavefunction amplitude.
Here, the graph IPR values also vary erratically from mode to mode (Fig. \ref{fig:fig4}(b)), but lie well below the maximum localization limit ($IPR_{IIb}$ = 1) and closer to the RMT prediction limit ($IPR_{IIb} = 1/B = 0.07$) \cite{Hul2009}.

The conclusions we draw from the above two methods are not exactly the same.
For Method IIa, two exemplary eigenfunction profiles are shown in Figs. \ref{fig:fig4}(c) and d, which correspond to the RMT and ergodic limits, respectively.
One may directly spot the different nature of these two modes based on their eigenfunction patterns, which is a convenient feature of the photonic crystal graph system.
For Method IIb, we present the eigenfunction profile of two neighboring graph modes in Figs. \ref{fig:fig4}(e) and (f).
The eigenmode in Fig. \ref{fig:fig4}(e) has a larger value of $IPR_{IIb}$ and shows a strongly localized distribution.
That in Fig. \ref{fig:fig4}(f) has a small value of $IPR_{IIb}$ and is more evenly distributed over the bonds.
The average value of IPR over all the modes is more meaningful in this case, and we have $\langle IPR_{IIa}\rangle=1.39$ and $\langle IPR_{IIb}\rangle=0.10$ which are closer to the ergodic and RMT limits, respectively, than to the localization limit.
Previous work \cite{Kaplan2001,Hul2004} on IPR on quantum graphs shows the surprising results that larger graphs, with the number of vertices $>$ 10, tend to show strong deviations from RMT predictions, while smaller graphs show better agreement.
{We note that by computing the $IPR_{IIa}$ on single loops rather than the entire graph, one may identify the modes trapped inside these loops from $IPR_{IIa} \sim B_{sl}$, where $B_{sl}$ is the total bond number of the single loop.
This study is discussed in the Appendix.}

Taking into account the eigenfunction and IPR statistics, it would appear that PC graphs are close to the random plane wave condition for tetrahedral-like graphs, but clear systematic differences from RMT predictions remain.
The complex vertex scattering properties of right-angle bends and T-junctions, along with the $\omega(k)$ dispersion relation of PC defect waveguide modes, suggest that PC defect graphs may be very effective for future wave chaotic studies.

\section{V. Conclusion}

{To conclude, we have designed and simulated an alternative chaotic graph system with photonic crystal defect waveguides that show a dispersion relation that is different from that utilized in coaxial cable microwave graphs.}
We show that a series of statistical studies can be carried out on {an ensemble of} closed graphs, including nearest-neighbor spacing statistics and eigenfunction statistics studies.
Because both the graph bonds and nodes can be probed, one may better analyze the non-universal features of chaotic graphs using a PC system.
{It is possible to adjust the degree of wave localization by engineering the scattering properties of the PC waveguide junctions.
This property of PC waveguides may facilitate further studies of localization phenomena in graphs, for example, the emergence or suppression of trapped modes.}

\section{Acknowledgements}

This work was supported by ONR under Grant No. N000141912481, AFOSR COE Grant FA9550-15-1-0171, DARPA WARDEN Grant HR00112120021, and the Maryland Quantum Materials Center.

\appendix*
\section{Appendix A: Sub-graph IPR Analysis}

\begin{figure*}
\centering
\includegraphics[width=0.9\textwidth]{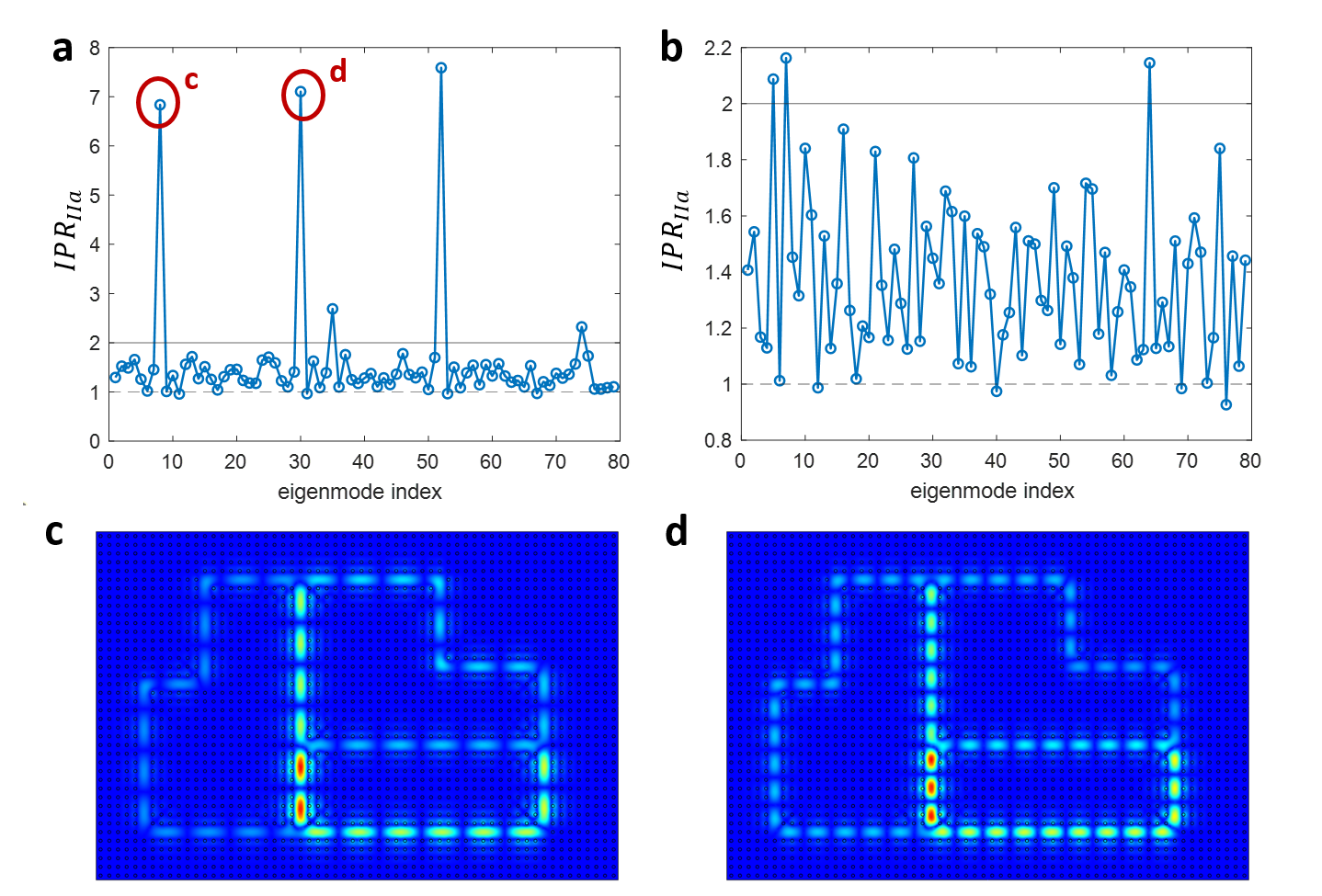}
\caption{\label{fig:sup1} 
\textbf{a.} The inverse participation ratio (IPR) of the left side of the graph, computed based on method IIa. The simulated $|E_z|$ profile of the circled modes in \textbf{a} is shown in \textbf{c} (\textbf{d}).
\textbf{b.} The IPR of the right side of the graph is computed based on method IIa.
The solid and dashed lines in \textbf{a.} and \textbf{b.} refer to the RMT and the maximum ergodic limits, respectively.
}
\end{figure*}

Here we give a brief discussion on applying the inverse participation ratio (IPR) analysis to only a sub-region of the graph structure.
Although the original definition of IPR is for the entire graph, we may apply the same computation in the context of subgraphs.
Here we conduct subgraph IPR studies with bonds that only belong to the left half (one single loop) or the right half (two loops) of the PC graph shown in Fig. \ref{fig:sup1}.
The left half loop IPR is shown in Fig. \ref{fig:sup1} (a), computed based on method IIa.
The IPRs computed with method IIb are not shown here.
The majority of the eigenmodes have IPRs between the RMT (IPR = 2) and the maximum ergodic limits (IPR = 1).
In the left half loop IPR test, we find three modes that have IPR values that reach the maximum localization limit (IPR = B, B = 7 for the left single loop).
We show the eigenfunction profile of the two circled modes in Fig. \ref{fig:sup1} (c) and (d).
These two modes indeed show a high degree of localization behavior, where the eigenfunction amplitude is mainly concentrated on one short bond.
Fig. \ref{fig:sup1} (b) shows the IPR analysis of the right half graph. 
Here the modes with a higher degree of localization do not stand out in this subgraph IPR analysis.
Thus we note that applying the IPR analysis to single loops can bring out the modes with a higher degree of localization.

\end{spacing}

\end{document}